\documentclass[journal,comsoc]{IEEEtran}
\usepackage[T1]{fontenc}
\usepackage{times}
\usepackage{soul}
\usepackage{url}
\usepackage[hidelinks]{hyperref}
\usepackage[utf8]{inputenc}
\usepackage[small]{caption}
\usepackage{graphicx}
\usepackage{amsmath}
\usepackage{amsthm}
\usepackage{booktabs}
\usepackage{algorithm}
\usepackage{algorithmic}
\usepackage{multirow}
\usepackage{threeparttable}
\usepackage{amsfonts}
\usepackage{float}
\usepackage{url}
\usepackage{bm}
\usepackage{xcolor}
\usepackage{subcaption}
\usepackage{amssymb}
\usepackage{makecell}
\usepackage{bbding}

\newcommand{\cmmnt}[1]{}

\ifCLASSINFOpdf
\else
\fi
\usepackage{amsmath}
\usepackage[cmintegrals]{newtxmath}
\begin{document}
%
\title{METTS: Multilingual Emotional Text-to-Speech by Cross-speaker and Cross-lingual Emotion Transfer}
\author{Xinfa Zhu,
        Yi Lei,
        Tao Li,
        Yongmao Zhang,
        Hongbin Zhou,
        Heng Lu,
        Lei Xie,~\IEEEmembership{Senior Member,~IEEE}
 \thanks{Corresponding author: Lei Xie}

\thanks{Xinfa Zhu, Yi Lei, Tao Li, Yongmao Zhang, and Lei Xie are with Audio, Speech and Language Processing Group (ASLP@NPU), the School of Computer Science, Northwestern Polytechnical University, Xi’an 710072, China. Email: xfzhu@mail.nwpu.edu.cn (Xinfa Zhu), leiyi@npu-aslp.org (Yi Lei), taoli@npu-aslp.org (Tao Li), zym@mail.nwpu.edu.cn (Yongmao Zhang), lxie@nwpu.edu.cn (Lei Xie)}

\thanks{Hongbin Zhou and Heng Lu are with Ximalaya Inc., Shanghai, China. Email: hongbin.zhou@ximalaya.com (Hongbin Zhou), bear.lu@ximalaya.com (Heng Lu)}
    }


%

%

\markboth{Journal of \LaTeX\ Class Files,~Vol.~14, No.~8, August~2015}%
{Wang \MakeLowercase{\textit{et al.}}: Bare Demo of IEEEtran.cls for IEEE Communications Society Journals}
%



\maketitle

\begin{abstract}

Previous multilingual text-to-speech (TTS) approaches have considered leveraging monolingual speaker data to enable cross-lingual speech synthesis. However, such data-efficient approaches have ignored synthesizing emotional aspects of speech due to the challenges of cross-speaker cross-lingual emotion transfer -- the heavy entanglement of \textit{speaker timbre}, \textit{emotion} and \textit{language} factors in the speech signal will make a system to produce cross-lingual synthetic speech with an undesired foreign accent and weak emotion expressiveness. This paper proposes Multilingual Emotional TTS (METTS) model to mitigate these problems, realizing both cross-speaker and cross-lingual emotion transfer. Specifically, METTS takes DelightfulTTS as the backbone model and proposes the following designs. First, to alleviate the foreign accent problem, METTS introduces~\textit{multi-scale emotion modeling} to disentangle speech prosody into coarse-grained and fine-grained scales, producing language-agnostic and language-specific emotion representations, respectively. Second, as a pre-processing step, formant shift based \textit{information perturbation} is applied to the reference signal for better disentanglement of speaker timbre in the speech. Third, a vector quantization based \textit{emotion matcher} is designed for reference selection, leading to decent naturalness and emotion diversity in cross-lingual synthetic speech. Experiments demonstrate the good design of METTS.
\end{abstract}


%
\IEEEpeerreviewmaketitle

\section{Introduction}
\label{sc:Introduction}

\IEEEPARstart{R}{ecent} years have witnessed significant progress in the quality and naturalness of synthetic speech thanks to the advances of neural text-to-speech (TTS) systems~\cite{DBLP:conf/interspeech/WangSSWWJYXCBLA17,DBLP:conf/nips/RenRTQZZL19,DBLP:conf/aaai/Li0LZL19,DBLP:conf/iclr/Chen0LLQZL21,DBLP:conf/interspeech/JeongKCCK21,DBLP:conf/icassp/WeissSBMK21,DBLP:conf/icml/KimKS21,DBLP:journals/taslp/LeiYWX22}. As near human parity performance has been reported in some closed TTS domains~\cite{DBLP:conf/acl/0006TQZL22}, \textit{diversity} and \textit{controllability} have become the new chasing target, including multi-speaker~\cite{DBLP:conf/icassp/WuSLW22}, multi-lingual~\cite{DBLP:conf/interspeech/ZhangWZWCSJRR19}, multi-emotion~\cite{DBLP:journals/taslp/LeiYWX22} as well as multi-style~\cite{DBLP:conf/icml/MinLYH21} scenarios. At the same time, \textit{data efficiency} is also highly desired as modern corpus-based TTS heavily relies on high-quality annotated data. This induces a variety of approaches better leveraging limited, low-resource as well as low-quality data~\cite{DBLP:conf/icassp/LiuYSY22,DBLP:conf/icassp/Yan0LQZSL21,DBLP:conf/interspeech/SaekiTY22}. 

This paper addresses an extremely diverse and controllable speech generation scenario -- multilingual emotional text-to-speech (METTS), particularly considering data efficiency by cross-speaker and cross-lingual emotion transfer. Specifically, METTS can produce bilingual emotional speech for each speaker after system building, while in the training data, each speaker speaks only one language (monolingual speaker), and some speakers have only neutral speech. Importantly, with only the neutral native speech (L1) data for a target speaker during training, helped with another emotional speaker in the target language (L2), our METTS system is able to produce the target speaker's emotional speech in the target language (L2) with reasonable proficiency and naturalness. METTS has significant applications such as foreign movie dubbing and computer-assisted language learning (CALL). However, building METTS is not a trivial task with three challenges.

\begin{itemize}
  \item

\textit{Foreign accent problem}. Different languages have quite different prosody patterns in pronunciation. In a typical cross-lingual TTS system, the accent of the source language may be inevitably delivered to the speech in the target language~\cite{DBLP:conf/interspeech/ZhangWZWCSJRR19}, leading to non-native synthetic speech with a strong foreign accent. This problem is prominent for cross-speaker and cross-lingual emotion transfer because emotional expressions are heavily reflected in prosody patterns through changes in pitch, loudness, speech rate and pauses.

\item

 \textit{Speech entanglement problem}. 
Only partial speakers in the training set have emotional speech data, while we need to enable each speaker in the training set generates emotional speech. The speaker timbre and emotion entanglement in speech may lead to \textit{speaker timbre leakage} when performing cross-speaker emotion transfer~\cite{DBLP:journals/spl/LeiYZXS22}. In other words, the source speaker's timbre may be inevitably transferred to the speech of the target speaker, making the synthetic speech of the target speaker sounds like the source speaker. Particularly, this problem becomes more severe for cross-speaker cross-lingual emotion transfer as the entanglement of three factors in speech -- speaker timbre, language and emotion complicates the disentanglement process.
\item
 \textit{Emotional diversity problem}. In order to synthesize diverse emotions, the TTS model usually needs additional emotional information as prior, such as an emotion ID or a reference speech sample. Compared to emotion ID, providing emotional representation through a reference encoder is apparently more diverse~\cite{DBLP:journals/corr/abs-2206-14866} as different references lead to different fine-grained emotion deliveries. However, such diversity raises a problem -- how to select an appropriate reference signal to exactly match the textual content~\cite{DBLP:conf/interspeech/MengL0LSXSZM22}. In this paper, cross-lingual reference selection, i.e., selecting a reference in L1 to match the text in L2, is a brand new problem for cross-speaker emotion transfer in the multilingual TTS system.
 
\end{itemize}

To address the above challenges, we propose METTS, a novel approach to synthesizing bilingual emotional speech for each monolingual speaker, even though some speakers do not have emotional speech data during model training. METTS is based on DeligtfulTTS~\cite{DBLP:journals/corr/abs-2110-12612}, a state-of-the-art non-autoregressive text-to-speech approach with improved Conformer~\cite{DBLP:conf/interspeech/GulatiQCPZYHWZW20} blocks to model the variation of speech prosody. Based on the skeleton of DelightfulTTS, our METTS leverages the following designs to achieve: 1) emotion transfer from a reference mel-spectrogram to synthesize speech (METTS-REF) and 2) automatically matching the most suitable reference embedding according to the input text and emotion ID to synthesize speech (METTS-ID).

First, we introduce \textit{multi-scale emotion modeling} to address the foreign accent problem. 
We believe emotion expressions in multilingual speech can be generally factorized into the shared similar prosody pattern~\cite{DBLP:conf/chi/DaiFM09, DBLP:conf/interspeech/SchullerMLR05} (e.g., high pitch for angry and low pitch for sad) conveyed in both languages and distinct fine details of prosody~\cite{DBLP:journals/ijst/KottiP12, DBLP:conf/mldm/FersiniMAA09} due to the different manners of pronunciation. 
This inspires us to model emotion with \textit{coarse} and \textit{fine} scales to respectively represent \textit{language-agnostic} and \textit{language-specific} emotion aspects in speech. To be specific, the coarse-grained emotion is modeled by a Global Style Token (GST)~\cite{DBLP:conf/icml/WangSZRBSXJRS18} layer with semi-supervised constraints to make the coarse-grained emotional representation language-agnostic; a Conditional Variational Autoencoder (CVAE) layer establishes a language-specific emotion representation by investigating the fine-grained prosody variation of speech, with the condition of language-dependent text input and the above coarse-grained emotional representation. In this way, METTS manages to successfully transfer emotion across languages via the coarse-grained language-agnostic emotional representation and produces native pronunciation without a foreign accent in emotion delivery through the fine-grained language-specific emotion representation. 

Second, we employ \textit{information perturbation}~\cite{DBLP:journals/spl/LeiYZXS22} to further address the issue of speech entanglement. Specifically, the reference speech is perturbed by randomly shifting its formant frequency, which allows the multi-scale emotion modeling module to generate a speaker-independent emotional representation. This speaker-independent signal can decouple the speaker timbre from reference speech in nature.

Third, to address the emotional diversity problem and select
an appropriate reference signal, we propose a vector quantization (VQ) based \textit{emotion matching} module. Instead of directly modeling the complicated regression between bilingual textual representation and emotional representation, we first use VQ to quantify coarse-grained language-agnostic emotion representation to form a reference pool and subsequently adopt an emotion matcher to match the bilingual textual representation with the reference pool. This allows METTS to realize \textit{reference-free} inference and produce more diverse emotional speech with reasonable expressiveness for both L1 and L2 text inputs.

The proposed METTS system is extensively evaluated on a demanding Mandarin-English bilingual TTS task. This task poses significant challenges due to substantial differences in pronunciation between Chinese and English, such as variations in syllable structure, tones, vowels, and consonant inventory, the absence of retroflex sounds in Chinese, and the presence of long vowels in English~\cite{han2013pronunciation,li2016contrastive}. Experimental results show that although the performance of intra-lingual emotional speech synthesis is better than that of more challenging cross-lingual emotional speech synthesis, METTS can produce bilingual emotional speech for each target speaker and effectively improve speech naturalness, speaker
similarity, and emotion similarity compared to other competitive methods. Furthermore, the component analysis validates the good design of our proposed model. Audio samples can be found on our demo page \footnote{https://anonymous-rep0.github.io/METTS/}.


The remainder of this paper is organized as follows. Section~\ref{sc:related work} provides a comprehensive review of related work in the field. Section~\ref{sc:method} presents a detailed description of the proposed approach. The experimental setups and results are described in Section~\ref{sc:experiments} and Section~\ref{sc:results}, respectively, where we analyze the performance of METTS and present the evaluation outcomes. In Section~\ref{component}, we delve into the component analysis, examining the contributions of each individual module in our system. Finally, Section~\ref{sc:conclusion} concludes the paper, summarizing the findings and highlighting future research directions.

\section{Related work}
\label{sc:related work}
Multi-lingual speech synthesis and emotional speech synthesis are two popular topics in the literature where transfer learning approaches -- transferring speaker voices across languages or transferring style/emotion across speakers -- are mainstream approaches better leveraging limited data. However, to the best of our knowledge, the current studies have not yet addressed both cross-speaker and cross-lingual emotion transfer. This is a more challenging task, as discussed in Section 1. Moreover, to improve the diversity of synthetic speech, cross-lingual reference selection is another new problem in reference-based multi-lingual emotional TTS. Therefore, here we review the prior arts in multilingual speech synthesis, emotional speech synthesis, and reference speech selection, respectively.

\subsection{Multilingual speech synthesis}



Developing a robust multilingual text-to-speech (TTS) system requires a unified representation of textual input. This can be achieved by merging phoneme sets from different languages or utilizing the International Phonetic Alphabet (IPA) to represent speech sounds. Based on the unified representation of textual input, several studies have explored a more efficient method that utilizes a single acoustic model with shared parameters across languages as an alternative to training separate models for each language~\cite{DBLP:conf/ssw/SitaramRRB16,DBLP:conf/interspeech/XueSXXW19,DBLP:conf/icassp/NachmaniW19}. Additionally, incorporating explicit language identification (ID) enables better control over language-specific prosody and improves the naturalness of synthetic speech~\cite{DBLP:conf/interspeech/LiZ16, DBLP:conf/interspeech/PengL22}.

However, due to distinct prosody patterns in different languages, 
cross-lingual synthetic audio often suffers from an undesired foreign accent problem in multilingual TTS systems. To address this challenge, some researchers investigate obtaining language-specific and speaker-independent prosodic representations through implicit disentanglement~\cite{DBLP:conf/interspeech/XinSTKS20,DBLP:conf/icassp/XinKTS21}. Typically, domain adversarial training strategies have been employed to encourage the model to learn disentangled representations of text and speaker identity~\cite{DBLP:conf/interspeech/ZhangWZWCSJRR19,DBLP:conf/interspeech/NekvindaD20,DBLP:conf/interspeech/PengL22}, where a gradient reversal layer is inserted before a speaker classifier. Furthermore, some studies~\cite{DBLP:conf/interspeech/ShangHZZ021,DBLP:conf/interspeech/RattcliffeWMKMC22} propose to use a style VAE encoder to alleviate the foreign accent problem by leveraging existing authentic styles during inference. Moreover, the triplet training scheme is utilized to overcome the accent problem by combining unseen speakers and language through fine-tuning~\cite{DBLP:conf/icassp/YeZSHRLL22}. CrossSpeech~\cite{DBLP:journals/corr/abs-2302-14370} obtains disentangled speaker and language representations through the speaker-independent generator and speaker-dependent generator.

It is important to note that while these approaches have shown promise in improving the performance of multilingual speech synthesis, challenges still remain in achieving emotionally expressive and foreign accent-free speech across different languages. 

\subsection{Emotional speech synthesis}

Significant progress has been made in the field of emotional speech synthesis in recent years~\cite{DBLP:journals/corr/abs-2206-14866,DBLP:journals/taslp/LeiYWX22,DBLP:journals/taslp/LiWXWX22,DBLP:journals/corr/abs-2110-04153}. When categorized emotion data is available for the target speaker, a straightforward approach is to model emotional expressions based on discrete emotion IDs~\cite{DBLP:conf/apsipa/AnLD17,DBLP:journals/corr/abs-1711-05447}. However, the resulting synthetic speech often exhibits over-averaged emotional expressions due to the discrete nature of emotion ID control. On the other hand, transfer learning approaches have been proposed, where a reference encoder is used to extract emotional representations from a reference signal, providing guidance for emotion synthesis. These emotion transfer approaches offer more flexibility and diversify the generated emotional speech~\cite{DBLP:conf/icml/Skerry-RyanBXWS18,DBLP:conf/interspeech/AkuzawaIM18,DBLP:conf/iscslp/LiYXX21}.

Emotion transfer is effective for generating emotional speech for speakers who only have neutral data, while it often faces a trade-off between speaker similarity and emotional expressiveness in synthetic speech. This trade-off results in either low speaker similarity or poor emotion similarity. To address this issue, disentangling these speech attributes and modeling them separately becomes necessary. Various schemes have been proposed for disentanglement. Some work implicitly decouples speech attributes in latent representations~\cite{DBLP:conf/interspeech/WangDYCLM21,DBLP:conf/interspeech/WangLZ0KM21}. Whitehill et al.~\cite{DBLP:conf/interspeech/WhitehillMMS20} use an unpaired training strategy and adversarial
cycle consistency scheme to disentangle emotion and speaker.  Li et al.~\cite{DBLP:journals/taslp/LiWXWX22} propose an emotion-disentangling module, which learns speaker-independence emotion embedding via an orthogonal loss with the speaker embedding. On the other hand, some studies~\cite{DBLP:conf/nips/ChoiLKLHL21,DBLP:conf/icassp/ChanQZH22} explicitly decouple speech attributes through bottleneck features, information perturbation, and other methods. Li et al.~\cite{DBLP:journals/spl/LeiYZXS22} learn emotion-related mel-spectrogram and speaker-related mel-spectrogram through information perturbation and generate emotionally expressive speech.

Considering human emotional expression's diverse and complex patterns, some research proposes to model emotional speech at multiple scales to capture rich emotional variations.~\cite{DBLP:journals/taslp/LeiYWX22,DBLP:journals/corr/abs-2205-07211,DBLP:journals/corr/abs-2210-15834}. For instance, the approach in~\cite{DBLP:journals/taslp/LeiYWX22} combines global emotion representation with local emotion representation at a fine-grained level, such as phoneme- or syllable-level, resulting in more natural and expressive speech. However, in the context of multilingual emotional speech synthesis, the entanglement between emotion and language becomes more complex and requires adequate consideration.

\subsection{Reference speech selection}

The selection of appropriate reference speech plays a crucial role in ensuring the naturalness and diversity of emotional expression in emotion transfer-based TTS approaches. In practical non-parallel transfer applications, the textual contents of reference audio are different from that of generated speech during inference, while they remain the same during training. This mismatch can result in degraded speech naturalness~\cite{DBLP:conf/icml/ChangSKZT22,DBLP:journals/corr/abs-2303-04289} and inappropriate emotional expressions. Since there is no text-matched reference available during inference, the problem of selecting suitable reference audio becomes challenging.

One straightforward approach to address the mismatch problem is leveraging multiple references during training and inference, such as through simple averaging embeddings of multiple references~\cite {DBLP:conf/icassp/GongWLZD22}. Recently, more sophisticated approaches based on context have been proposed for reference selection. For instance, ProsodySpeech~\cite{DBLP:conf/icassp/YiHPWX22} utilizes a Prosody Distributor that employs an attention mechanism to select references at the phone level. Inspired by Contrastive Language-Image Pre-training (CLIP)~\cite{DBLP:conf/icml/RadfordKHRGASAM21}, CALM~\cite{DBLP:conf/interspeech/MengL0LSXSZM22} incorporates a Contrastive Acoustic-Linguistic Module to select reference speeches based on the input text.

However, as human emotions and the contextual content of multilingual scripts are highly complicated, addressing emotional reference selection and accounting for the diverse contextual content in multilingual scenarios remain essential research topics.

\section{Methodology}
\label{sc:method}

This section first gives an overview of METTS, followed by the motivation and design of each module. The training pipeline will also be introduced in this section.

\begin{figure*}[htb]
\centering
\begin{minipage}[b]{0.8\linewidth}
    \centerline
  \centerline{\includegraphics[width=\textwidth]{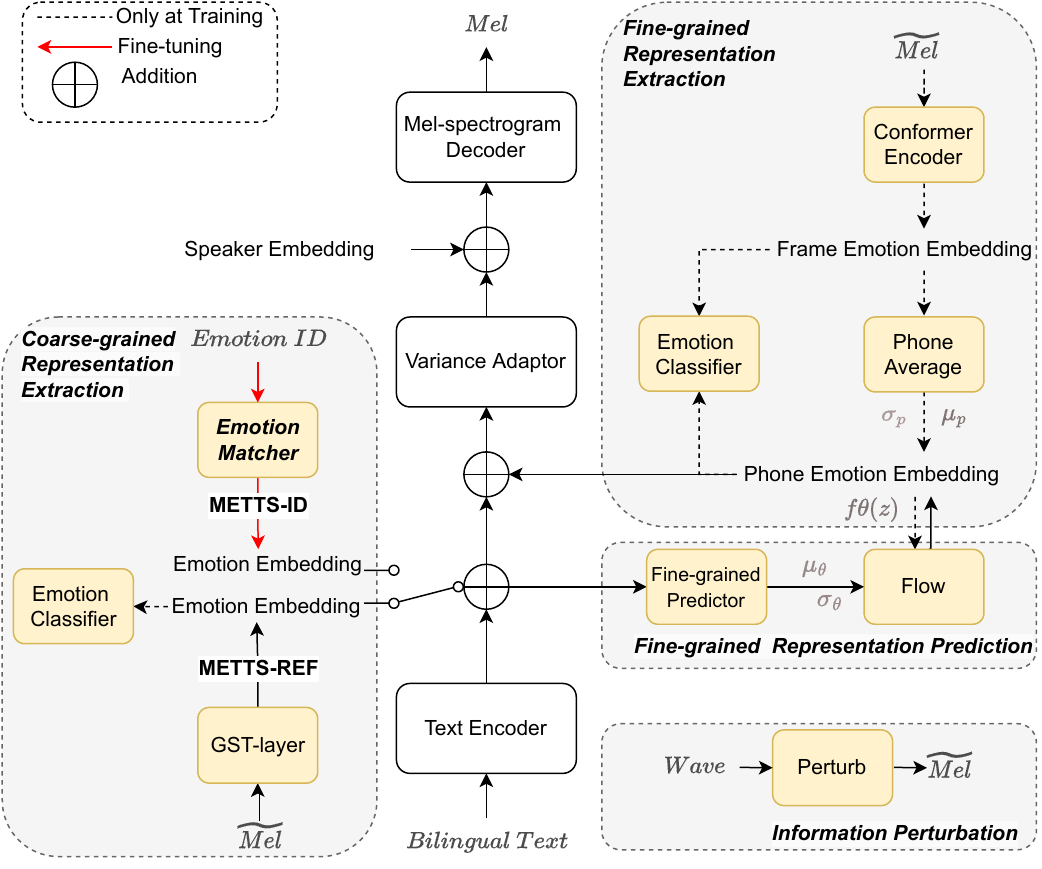}}
\end{minipage}
\caption{The architecture of METTS.}
\label{fig_1}
\end{figure*}

\subsection{Overview}
\label{sc:Overview}

The proposed framework is built on a multilingual text processing front-end, which supports both Chinese and English. The front end encodes Chinese text input into labels of phonemes, tones, word boundaries, and prosodic boundaries while encoding English text input into labels of phonemes. Note that the Chinese phoneme set is based on Pinyin, while English is based on CMU-Dict. Therefore, we merge the Chinese and English phoneme sets for unified textual inputs of the bilingual TTS system.

As shown in Figure~\ref{fig_1}, the backbone of METTS is based on DelightfulTTS~\cite{DBLP:journals/corr/abs-2110-12612}, which consists of a text encoder, a variance adaptor, and a mel-spectrogram decoder. Notably, the improved Conformer~\cite{DBLP:conf/interspeech/GulatiQCPZYHWZW20} structure better model local and global dependency of mel-spectrogram, which has led DelightfulTTS to win the Blizzard Challenge 2021.
In general, METTS updates DelightfulTTS with \textit{multi-scale emotion modeling} to achieve natural multilingual emotional TTS by both emotion reference (METTS-REF) and ID (METTS-ID) as the control signal. Specifically, the coarse-grained emotion embedding is provided by a GST layer or an emotion matcher, while the fine-grained emotion embedding is obtained from the CVAE module. 
Moreover, a perturb module is introduced to distort the speaker timbre of the reference signal for decoupling timbre from speech. 
The speaker embedding is obtained through a lookup table and is added to the output of the variance adaptor, which is then fed into the mel-spectrogram decoder to synthesize the final mel-spectrogram.

\subsection{Multi-scale emotion modeling}
\label{sc:Multi-scale emotion modeling}

In general, the emotional expressions of multi-lingual speech could be factorized to the shared prosody pattern and distinct fine details of prosody due to the different manners of pronunciation. METTS utilizes the Global Style Tokens (GST) and Conditional Variational Autoencoder (CVAE) modules to establish coarse-grained language-agnostic and fine-grained language-specific emotional representations. 

The GST module employs a reference encoder to encode mel-spectrogram into a hidden representation and utilizes multi-head attention to calculate the global emotional style tokens. Notably, L2 normalization is applied to the global emotional representation, eliminating magnitude-related information that may vary across languages and speakers. This normalization improves the generalization ability of the model, allowing for emotion embedding control based solely on angular information geometrically~\cite{DBLP:conf/smc/KimLLJL21}. By mapping the emotional representation of different languages to the same global tokens, the GST module forms the language-agnostic representation.

The CVAE module focuses on learning fine-grained emotion expression from the mel-spectrogram, with text and coarse-grained emotion conditions. It utilizes a conformer block and a GRU layer to extract frame-level emotion embeddings, which are then downsampled to the phoneme level based on duration. These phoneme-level embeddings are used to derive the mean and variance of the distribution of phoneme-level prosody. Additionally, taking inspiration from VITS~\cite{DBLP:conf/icml/KimKS21}, a fine-grained predictor is designed to predict the distribution of fine-grained emotion from text. To improve the expressiveness of the predicted distribution, a normalizing flow technique is employed, enabling an invertible transformation from a simple distribution to a more complex one. By learning fine-grained emotional representations that are consistent with the text, the CVAE module forms the language-specific representation.

To ensure that the extracted multi-scale representations are relevant to emotions, even for training data without emotion annotations, a semi-supervised strategy is employed, which includes an emotion classifier. Specifically, only the embeddings of annotated audio are used to supervise the emotion classifier for both coarse- and fine-grained representations. The audio without emotion annotations is not involved to optimize the emotion classifiers and is utilized to train the acoustic model by the extracted embeddings. Furthermore, the frame and phoneme emotion embeddings are processed through a GRU layer to extract a single vector, which is then used as input to the emotion classifier.

\subsection{Speaker disentanglement based on information perturbation}

In our multilingual emotional speech synthesis setup, where each speaker is mono-lingual and only some speakers in the training set have emotional speech data, the entanglement of speaker timbre with emotion and language poses a challenge, resulting in synthetic speech with low speaker similarity and unusual emotional expression and pronunciation.

To address this issue more comprehensively, we adopt a pre-processing step that utilizes a signal perturbation module to remove the speaker timbre information. Specifically, we apply a dynamic \textit{formant} shift to the mel-spectrogram of the reference speech. Speech formants are primarily determined by the size, shape, and position of the vocal tract, which are highly specific to each speaker and represent their vocal identity~\cite{DBLP:journals/taslp/YooLY15}. By performing a formant shift function, denoted as $fs$, on the original waveform $Wave$, we obtain a speaker-independent signal denoted as $\widetilde{Wave} = fs (Wave)$.

Subsequently, we extract the mel-spectrogram of the perturbed wave, denoted as $\widetilde{Mel}$, which serves as the input for emotion representations extraction. The perturbation module perturbs the timbre of the recordings at a random scale by each step during training, allowing the GST and CVAE modules to learn a speaker-independent representation and effectively disentangle the speaker's timbre from speech.

\begin{figure*}[htb]

\begin{minipage}[b]{\linewidth}
    \centering
    \includegraphics[width=0.7\textwidth]{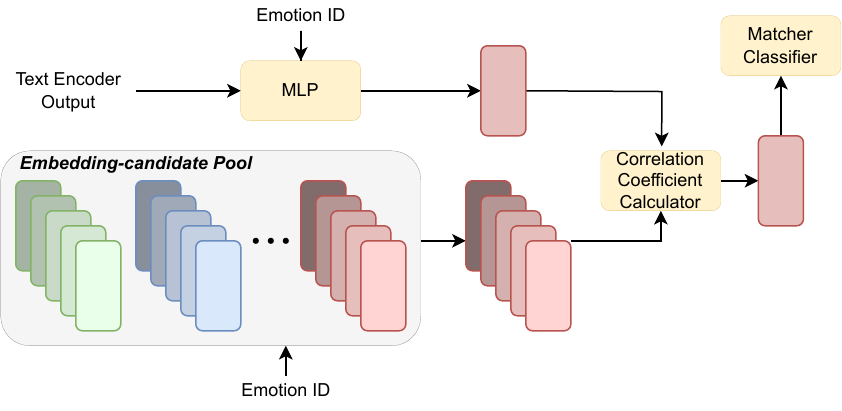}
\end{minipage}
\caption{The architecture of emotion matcher.}
\label{fig_2}

\end{figure*}

\subsection{VQ based emotion matcher}

During inference, different reference usually lead to different emotional expression of the synthetic speech. The selection of an appropriate reference is crucial for achieving natural speech and accurate emotional expression. Therefore, we propose an emotion matcher that automatically matches the most suitable reference embedding based on the textual content. Figure~\ref{fig_2} illustrates the architecture of the emotion matcher, which takes the text encoder output and the language ID as input and generates the optimal reference embedding, facilitating reference-free inference in METTS.

Directly modeling the complex relationship between bilingual textual representation and emotional representation is quite challenging, so we employ VQ to quantize the coarse-grained emotion representation, forming a reference pool. Subsequently, we predict the correct codebook from the reference pool for the current utterance. This transforms the intricate regression task into a simpler classifier task, simplifying the modeling process.

Specifically, in our approach, we begin by extracting the coarse-grained emotional representation and predicted emotion ID for all training audio samples using the GST layer and the emotion classifier. Subsequently, we apply VQ to quantize the coarse-grained emotional representation. To achieve this, we employ the k-means algorithm to obtain $N$ clusters for each emotion category, resulting in a total of $N \times M$ reference embeddings ($M$ is the number of emotion categories), which form the reference pool.

To select the appropriate reference from the pool, the emotion matcher utilizes a multilayer perceptron (MLP) to generate a text-emotion vector. This vector captures the contextual information related to emotion by taking the text encoder output and emotion ID as inputs. Using the text-emotion vector and the embedding-candidate pool, we calculate the correlation coefficient (CC) matrix between them. The calculation of CC is similar to the process of Scaled Dot-Product Attention~\cite{DBLP:conf/nips/VaswaniSPUJGKP17} and is defined as:

\begin{equation}
CC(V_{t},E_{c}) = \text{softmax}\left(\frac{V_{t}E_{c}^T}{\sqrt{d_{E_C}}}\right),
\end{equation}
where $V_t$ and $E_c$ represent the text-emotion vector and embedding candidates, respectively, and $d_{E_C}$ denotes the dimension of the embedding candidates. The softmax function is applied to normalize the correlation coefficients. The embedding with the highest correlation coefficient is selected as the coarse-grained emotion representation for the TTS system.

To ensure that the selected embedding corresponds to the input text, a matcher classifier is introduced to supervise the CC matrix. It ensures that the embedding with the highest correlation coefficient is the cluster center corresponding to the input text.

\subsection{Training and fine-tuning}
For flexible control of the generated emotional expressions, we use pre-training and fine-tuning procedures to conduct multilingual emotional speech synthesis from a reference signal and manual emotion ID, respectively.

The training objective of METTS-REF is
\begin{equation}
\begin{aligned}  
\mathcal{L}_{\mathrm{pretrain}} = & 0.05 * \mathcal{L}_{\mathrm{kl}} +  \mathcal{L}_{\mathrm{prosody}} + \\
& 0.1 * \mathcal{L}_{\mathrm{emo}} + \mathcal{L}_{\mathrm{ssim}}+  \mathcal{L}_{\mathrm{iter}},
\end{aligned}\label{eq:eq1}
\end{equation}
where $\mathcal{L}_{\mathrm{prosody}}$ is the L1 loss between the predicted pitch/energy/duration and the ground-truth pitch/energy/duration, $\mathcal{L}_{\mathrm{emo}}$ is the semi-supervised crossentropy loss of emotion classifier. $\mathcal{L}_{\mathrm{kl}}$ is the KL divergence to predict the phoneme-level emotion distribution from text encoder output, and $\mathcal{L}_{\mathrm{iter}}$ is the sum of mel-spectrogram L1 loss between the predicted and ground-truth mel-spectrogram in each Conformer block. Moreover, we use structural similarity $\mathcal{L}_{\mathrm{ssim}}$ ~\cite{DBLP:journals/tip/WangBSS04} to measure the similarity between predicted and ground-truth mel-spectrogram in the final Conformer block.

The purpose of fine-tuning is to support METTS-ID. 
During fine-tuning, we use the ground-truth clustering center as the coarse-grained emotion representation for the TTS model and jointly optimize the emotion matcher. To stabilize the joint training, we freeze the GST-layer and emotion classifier as a discriminator to distinguish the emotion category of the generated mel-spectrogram.

The fine-tuning objective is
\begin{equation}   
\mathcal{L}_{\mathrm{finetune}} = \mathcal{L}_{\mathrm{match}} + \mathcal{L}_{\mathrm{disc}} + \mathcal{L}_{\mathrm{base^{'}}},
\end{equation}
where $\mathcal{L}_{\mathrm{match}}$ is the cross entropy loss between the selected clustering centre and the actual clustering centre in the emotion matcher, $\mathcal{L}_{\mathrm{disc}}$ is the emotion classification loss of the predicted mel-spectrogram taking GST layer as an discriminator, and $\mathcal{L}_{\mathrm{base^{'}}}$ means removing $\mathcal{L}_{\mathrm{emo}}$ from the pre-trained model objectives. 

\begin{table*}[h]
\centering
\caption{Dataset for the multilingual emotional TTS.}
\setlength{\tabcolsep}{3.0mm}
\label{tab:data}
\begin{tabular}{c|c|ccccccc|c}
\toprule
\multirow{2}{*}{Corpus}  & \multirow{2}{*}{Language} & \multicolumn{7}{c|}{Emotion (sentences)}   & \multirow{2}{*}{ Usage} \\
 &     & Neutral & Happy  & Surprise &Sadness & Angry & Disgust & Fear &   \\ \midrule
CN1        & Chinese  &5k    &0.5k   &0.5k     &0.5k    &0.5k   &0.5k   &0.5k       &Training\&Evaluation               \\
CN1     & Chinese  &5k   &2k    &2k    &2k   &2k  &2k  &2k     &Training\&Evaluation   \\
EN1       & English  &10k   &-   &-     &-    &-   &-   &-        &Training\&Evaluation   \\ 
EN2      & English  &10k   &-    &-     &-    &-   &-   &-       &Training\&Evaluation   \\ \bottomrule
\end{tabular}
\end{table*}

\section{Experimental Setups}
\label{sc:experiments}

This section introduces the database configuration, training setups, compared methods, and evaluation methods.

\subsection{Dataset} 
\label{sc:database}

To assess the performance of METTS, we conduct a series of experiments on Chinese and English datasets, as shown in Table~\ref{tab:data}. These two languages have tremendous pronunciation differences, which poses a challenge for multilingual speech synthesis. The Chinese dataset includes audio clips from two female speakers, denoted as CN1 and CN2, expressing six types of emotions~(anger, fear, happiness, sadness, surprise, and neutral). The total number of audio clips was 22,205, approximately 21 hours of audio in sum. The English dataset includes audio clips from two female speakers, EN1 and EN2, for 19,676 audio clips, approximately 20 hours. There is no apparent emotional expression in English datasets. 
All data are studio-quality recorded at 48KHz.

\subsection{Training setups}
\label{exset}

For all texts, the TTS front end encodes Chinese text input into phonemes, tones, word boundaries, and prosodic boundaries while decoding English text input into labels of phonemes. We down-sample all the audios into 24k Hz and set the frame and hop sizes to 1200 and 300, respectively, when extracting optional auxiliary acoustic features like pitch and mel-spectrogram. The auxiliary pitch and energy contour is extracted through WORLD~\cite{DBLP:journals/ieicet/MoriseYO16}, and the implementation of formant shifting is achieved by Praat, following the NANSY~\cite{DBLP:conf/nips/ChoiLKLHL21} model. The phoneme duration is obtained through an HMM-based force alignment model~\cite{Sjlander2003AnHS}.

METTS takes DelightfulTTS~\cite{DBLP:journals/corr/abs-2110-12612} as the backbone, which consists of an encoder and decoder, both containing six conformer blocks. The dimensions of the emotion embedding and speaker embedding are set to 384. The CVAE module uses the Flow setting from VITS~\cite{DBLP:conf/icml/KimKS21}, and the dimension of the fine-grained emotion embedding is set to 16. The multi-layer perceptron (MLP) consists of one conformer block layer, six two-dimensional convolution layers, and one gated recurrent unit (GRU) layer, which outputs a 384-dimensional vector. The number of clusters in the k-means algorithm $N$ is set to 64. All classifiers have the same structure that consists of 3 fully connected layers with the Relu activation function.

All models are trained up to 400k steps on two 2080Ti GPUs with a batch size of 12 and use a MelGAN~\cite{DBLP:conf/nips/KumarKBGTSBBC19} vocoder to convert the generated mel-spectrogram into waveforms.

\subsection{Comparison methods}

As this work is the first attempt, to the best of our knowledge, to synthesize foreign emotional speech through emotion transfer from reference speech or directly based on emotion ID, there are no existing methods directly comparable to our proposed approach. However, we compare our proposed METTS with the most relevant and recent methods in the field to provide a fair evaluation.
To ensure fairness in the comparison, we implement the following comparison models on the delightful TTS model backbone and maintain identical training setups.

\begin{itemize}
  \item
  \textbf{CET}~\cite{DBLP:journals/corr/abs-2110-04153} is a powerful Cross-speaker Emotion Transfer speech synthesis system, which defines several emotion tokens that are trained to be highly correlated with corresponding emotions by a semi-supervised training strategy. Speaker condition layer normalization is implemented to eliminate the down-gradation to the timbre similarity for cross-speaker emotion transfer. During inference, the model transfers emotion from a reference mel-spectrogram to the synthetic speech.
  \item
  \textbf{M3}~\cite{DBLP:conf/interspeech/ShangHZZ021} is a Multi-speaker, Multi-style, and Multi-lingual text-to-speech system, which utilizes a speaker conditional variational encoder and conducts adversarial speaker training by the gradient reversal layer. Moreover, the model uses a Mixture Density Network (MDN) for mapping text and the extracted style vectors for each speaker. In inference time, the model predicts emotion representation according to emotion ID and text to synthesize speech.
  \item
  \textbf{METTS-REF} is the proposed model that transfers emotion from a reference mel-spectrogram to synthesize speech. 
  \item
  \textbf{METTS-ID} is the proposed model that automatically matches the most suitable reference embedding according to the input text and emotion ID to synthesize speech.
\end{itemize}

\begin{table*}[htb]
\centering
\caption{Results of subjective evaluation with 95$\%$ confidence interval for Chinese speakers.}
\label{tab_1}
\begin{tabular}{@{}c|ccc|ccc@{}}
\toprule
                             & \multicolumn{3}{c|}{Chinese Text}                            & \multicolumn{3}{c}{English Text}                      \\ \midrule
Model                        & Naturalness & Speaker Similarity & Emotion Similarity & Naturalness & Speaker Similarity & Emotion Similarity \\ \midrule
METTS-REF    & \textbf{4.11±0.12} & \textbf{3.94±0.16} & \textbf{4.12±0.14} & 4.00±0.11 & \textbf{3.94±0.12} & \textbf{3.44±0.22} \\
METTS-ID     & 4.07±0.13          & 3.88±0.16          & 3.95±0.13          & \textbf{4.06±0.18}          & 3.77±0.20          & 3.24±0.22          \\
\textbf{CET}~\cite{DBLP:journals/corr/abs-2110-04153}      & 3.65±0.14          & 3.69±0.12          & 4.00±0.11          & 3.01±0.19          & 3.35±0.14          & 3.39±0.16          \\
M3~\cite{DBLP:conf/interspeech/ShangHZZ021}     & 2.69±0.19          & 3.35±0.15          & 3.21±0.18          & 2.49±0.23          & 3.46±0.16          & 2.97±0.16          \\
\bottomrule
\end{tabular}
\end{table*}

\begin{table*}[htb]
\centering
\caption{Results of subjective evaluation with 95$\%$ confidence interval for English speakers.}
\label{tab_2}
\begin{tabular}{@{}c|ccc|ccc@{}}
\toprule
                             & \multicolumn{3}{c|}{Chinese Text}                            & \multicolumn{3}{c}{English Text}                      \\ \midrule
Model                        & Naturalness & Speaker Similarity & Emotion Similarity & Naturalness & Speaker Similarity & Emotion Similarity \\ \midrule
METTS-REF    & 3.91±0.14          & \textbf{3.68±0.18} & 3.71±0.15          & 3.95±0.14          & \textbf{3.82±0.16} & \textbf{3.44±0.19} \\
METTS-ID     & \textbf{4.02±0.15} & 3.57±21            & \textbf{3.73±0.17} & \textbf{4.05±0.18} & 3.74±0.18          & 3.26±0.14          \\
\textbf{CET}~\cite{DBLP:journals/corr/abs-2110-04153}      & 3.08±0.16          & 2.88±0.17          & 3.41±0.16          & 2.89±0.12          & 3.33±0.15          & 3.21±0.20          \\
M3~\cite{DBLP:conf/interspeech/ShangHZZ021}     & 2.72±0.15          & 3.17±0.15          & 3.01±0.18          & 2.41±0.19          & 3.24±01.5          & 2.81±0.18          \\
\bottomrule
\end{tabular}
\end{table*}

\subsection{Evaluation metrics} 
\label{sc:evaluation}


To evaluate the performance of the benchmark systems, we conduct a comprehensive set of evaluation methods. We prepare two test sets consisting of forty English texts and forty Chinese texts. For each speaker and emotion category, we generate samples, resulting in a total of 1,920 samples (2 languages $\times$ 40 texts $\times$ 4 speakers $\times$ 6 emotions) for evaluation. For models that transfer emotion from a reference mel-spectrogram, we provide randomly selected mel-spectrograms from CN\_spk1 as the reference. For models that synthesize speech based on emotion ID, we provide the corresponding emotion ID as input.

For subjective evaluation, we conducted two types of human perceptual rating experiments.  A total of twenty-two volunteers with basic English skills participate in these experiments. Mean Opinion Score (MOS)~\cite{shang2021incorporating} is used to evaluate the naturalness of the synthetic speech. Participants are asked to rate the speech on a scale ranging from 1 to 5, reflecting the influence of foreign accents and emotion on naturalness. The rating criteria are as follows: bad = 1, poor = 2, fair = 3, good = 4, great = 5, with 0.5-point increments. Similarity Mean Opinion Scores (SMOS)~\cite{Li2021ControllableCE} is adopted to subjectively evaluate the synthetic speech from two aspects: emotion similarity and speaker similarity. Participants are asked to rate the speech's similarity to a given emotional reference and the similarity to the reference of the target speaker. The rating scale and criteria are the same as those used in the MOS evaluation.

For objective evaluation, we measure speaker cosine similarity, character error rate (CER), and word error rate (WER) for the synthetic audio. To measure speaker cosine similarity, we train an ECAPA-TDNN~\cite{DBLP:conf/interspeech/DesplanquesTD20} model trained on 3,300 hours of Mandarin speech and 2,700 hours of English speech from 18,083 speakers to extract x-vectors. We extract the averaged x-vector of all utterances for each English speaker and six averaged x-vectors for each emotion category of the Chinese speaker. We then extract the x-vector of the synthetic audio and calculate the cosine distance. A higher cosine similarity indicates a more similar speaker timbre. To evaluate CER and WER, we use an open-source model provided by the WeNet community~\cite{DBLP:conf/interspeech/YaoWWZYYPCXL21}, which uses the U2++ conformer architecture and is trained on 10,000 hours of open-source Gigaspeech English data~\cite{DBLP:conf/interspeech/ChenCWDZWSPTZJK21} and 10,000 hours of open-source WeNet Mandarin data~\cite{DBLP:conf/icassp/ZhangLGSYXXBCZW22}, respectively. A higher CER or WER indicates less accurate pronunciation.

\section{Experimental results}
\label{sc:results}

This section evaluates the performance of each system to produce bilingual emotional speech for Chinese and English speakers. The comparison between METTS and other methods is presented and discussed.

\subsection{Subjective evaluation}

We initially conducts a subjective evaluation to assess the performance of the generated multilingual emotional speech in terms of speech naturalness, speaker similarity, and emotion similarity for both Chinese and English speakers. The evaluation results, as presented in Table~\ref{tab_1} and Table~\ref{tab_2}, demonstrate that the proposed METTS family consistently outperforms the baseline models across all evaluation metrics for both Chinese and English speakers. Notably, all models exhibit a performance degradation during cross-lingual emotional speech synthesis, indicating that the synthetic Chinese speech for English speakers generally has lower quality compared to that for Chinese speakers. Nevertheless, the proposed METTS family demonstrates relatively minor degradation in performance during cross-lingual emotional speech synthesis, suggesting its capability to generate natural and fluent foreign speech for a given target speaker.

Comparing the different models in the METTS family, METTS-REF achieves the highest speaker and emotion similarity scores, indicating its effectiveness in transferring emotions from reference to synthetic speech. On the other hand, METTS-ID achieves almost the highest naturalness score and comparable emotion similarity to METTS-REF. This result validates the efficacy of the emotion matcher module in accurately matching a suitable reference embedding to synthesize more natural speech. Furthermore, there are two exceptional cases worth mentioning. In Table~\ref{tab_1}, METTS-REF achieves the highest naturalness score in synthesizing Chinese emotional speech for Chinese speakers, indicating that intra-lingual emotion expressions of different speakers are similar. In Table~\ref{tab_2}, METTS-ID obtains the highest emotion similarity score in synthesizing Chinese emotional speech for English speakers, which suggests that the coarse-grained emotion embedding provided by the emotion matcher module is close to that of the emotion encoder for English speakers in this particular condition.

CET demonstrates similar emotion similarity to METTS-REF under specific test conditions, indicating its powerful ability in emotion transfer. However, CET is primarily designed for inter-language emotion transfer and relies on a single-scale emotion representation, which is hard to capture the diverse emotional expressions across different languages. As a result, the synthetic speech may exhibit a heavy accent. Therefore, CET receives lower scores in naturalness and speaker similarity evaluations. In contrast, our proposed METTS model incorporates multi-scale emotion modeling to capture both language-specific and language-agnostic emotional expressions, effectively avoiding the entanglement of accents with emotions.
Furthermore, M3 performs poorly across all evaluation metrics. M3 assumes a strong correlation between style coding and the speaker's attributes and content~\cite{DBLP:conf/interspeech/ShangHZZ021}, which leads to an entanglement between the speaker's timbre and emotion. Additionally, the domain adversarial training used in M3 for speaker timbre disentanglement is not stable~\cite{DBLP:conf/iclr/AcunaLZF22}. In contrast, our proposed model employs information perturbation to effectively remove the speaker's timbre, resulting in a more stable and practical approach.

\subsection{Objective evaluation}
\label{sc:emotrans}

To comprehensively evaluate the performance of our multilingual emotional TTS system, we conduct objective tests to measure speaker cosine similarity, character error rate (CER) for synthetic Chinese speech, and word error rate (WER) for synthetic English speech.

\begin{table}[htb!]
\centering
\caption{Results of objective evaluation for Chinese speakers.}
\label{tab_7}
\begin{tabular}{@{}c|cc|cc@{}}
\toprule
                             & \multicolumn{2}{c|}{Chinese Text} & \multicolumn{2}{c}{English Text} \\ \midrule
Model                        & Cosine Similarity                & CER          & Cosine Similarity             & WER            \\ \midrule
METTS-REF  & \textbf{0.813} & 0.48          & \textbf{0.753}  & 5.60                  \\
METTS-ID & 0.805          & 0.48          & 0.711  & \textbf{5.46} \\
\textbf{CET}~\cite{DBLP:journals/corr/abs-2110-04153}  & 0.726          & \textbf{0.35} & 0.638   & 12.65                 \\
M3~\cite{DBLP:conf/interspeech/ShangHZZ021}  & 0.754          & 11.02         & 0.673  & 55.32                 \\ \bottomrule
\end{tabular}
\end{table}

\begin{table}[htb!]
\centering
\caption{Results of objective evaluation for English speakers.}
\label{tab_8}
\begin{tabular}{@{}c|cc|cc@{}}
\toprule
                             & \multicolumn{2}{c|}{Chinese Text} & \multicolumn{2}{c}{English Text} \\ \midrule
Model                        & Cosine Similarity                & CER          & Cosine Similarity             & WER            \\ \midrule
METTS-REF   & \textbf{0.735}            & 1.38                           & 0.769 & 5.51                                              \\
METTS-ID    & 0.709            & 1.36                           & \textbf{0.786}  & \textbf{2.15}                           \\
\textbf{CET}~\cite{DBLP:journals/corr/abs-2110-04153}     & 0.659                     & \textbf{1.24}                  & 0.663                   & 8.05                           \\
M3~\cite{DBLP:conf/interspeech/ShangHZZ021}           & 0.671                     & 16.74                          & 0.704                       & 38.22                                      \\ \bottomrule
\end{tabular}
\end{table}


The objective test results presented in Tables~\ref{tab_7} and Table~\ref{tab_8} confirm the observations from the subjective evaluation, highlighting the distinction between inter-lingual and cross-lingual speech synthesis. The METTS family achieves the highest speaker cosine similarity, demonstrating the effectiveness of our approach in disentangling speaker timbre from both emotion and language. Furthermore, the METTS family achieves lower CER and WER, indicating its stability in generating intelligent, natural-sounding multilingual emotional speech.


It is worth noting that CET achieves the lowest Chinese CER, showcasing its ability in intra-lingual emotion transfer. However, the higher English WER of CET reflects the significant challenges of cross-lingual emotion transfer, which aligns with the subjective evaluation results concerning naturalness. Additionally, M3 fails to effectively address accent-related challenges in multilingual emotional speech synthesis, resulting in incorrect pronunciation and yielding the highest CER and WER. Furthermore, the results of speaker cosine similarity suggest that information perturbation for speaker timbre removal employed in our approach is more effective than the SALN method used in CET and the speaker adversarial training method in M3 in terms of speaker disentanglement in multilingual emotional speech synthesis.

\begin{table*}[htb]
\centering
\caption{Results of Ablation study with 95$\%$ confidence interval for Chinese speakers.}
\label{tab_9}
\begin{tabular}{@{}l|ccc|ccc@{}}
\toprule
                             & \multicolumn{3}{c|}{Chinese Text}                            & \multicolumn{3}{c}{English Text}                      \\ \midrule
Model                        & Naturalness & Speaker Similarity & Emotion Similarity & Naturalness & Speaker Similarity & Emotion Similarity \\ \midrule
METTS-REF     & \textbf{4.11±0.12} & \textbf{3.94±0.16} & 4.12±0.14 & \textbf{4.00±0.11}  & \textbf{3.94±0.12}  & 3.44±0.22 \\
\: - GST                 & 3.50±0.15 & 3.82±0.17 & 3.19±0.18  & 3.69±0.13 & 3.80±0.18  & 3.07±0.17 \\
\: - CVAE              & 3.95±0.12 & 3.81±0.16  & 3.88±0.12 & 3.43±0.14 & 3.82±0.13  & 3.39±0.17 \\
\: - Perturb & 4.02±0.12 & 3.87±0.15 &\textbf{4.19±0.14} & 3.45±0.17 & 3.55±0.16   & \textbf{3.58±0.21} \\ \bottomrule
\end{tabular}
\end{table*}

\begin{table*}[htb]
\centering
\caption{Results of Ablation study with 95$\%$ confidence interval for English speakers.}
\label{tab_10}
\begin{tabular}{@{}l|ccc|ccc@{}}
\toprule
                             & \multicolumn{3}{c|}{Chinese Text}                            & \multicolumn{3}{c}{English Text}                      \\ \midrule
Model                        & Naturalness & Speaker Similarity & Emotion Similarity & Naturalness & Speaker Similarity & Emotion Similarity \\ \midrule
METTS-REF     & \textbf{3.91±0.14} & \textbf{3.68±0.18} & 3.71±0.15 & \textbf{3.95±0.14}  & \textbf{3.82±0.16}  & 3.44±0.19 \\
\: - GST                & 3.75±0.19          & 3.52±0.21          & 3.24±0.21          & 3.73±0.19          & 3.64±0.21          & 3.17±0.18          \\
\: - CVAE             & 3.71±0.18          & 3.56±0.19          & 3.58±0.19          & 3.17±0.18          & 3.76±0.21          & 3.30±0.20          \\
\: - Perturb   & 3.85±0.21 & 3.19±0.27          & \textbf{3.82±0.19} & 3.50±0.22 & 3.26±0.29                  & \textbf{3.52±0.20} \\ \bottomrule
\end{tabular}
\end{table*}

\subsection{Visual analysis of emotional representation}

\begin{figure}[htb]
\begin{minipage}[b]{\linewidth}
\centering
\begin{subfigure}[b]{0.9\textwidth}
\centering
\includegraphics[width=\textwidth]{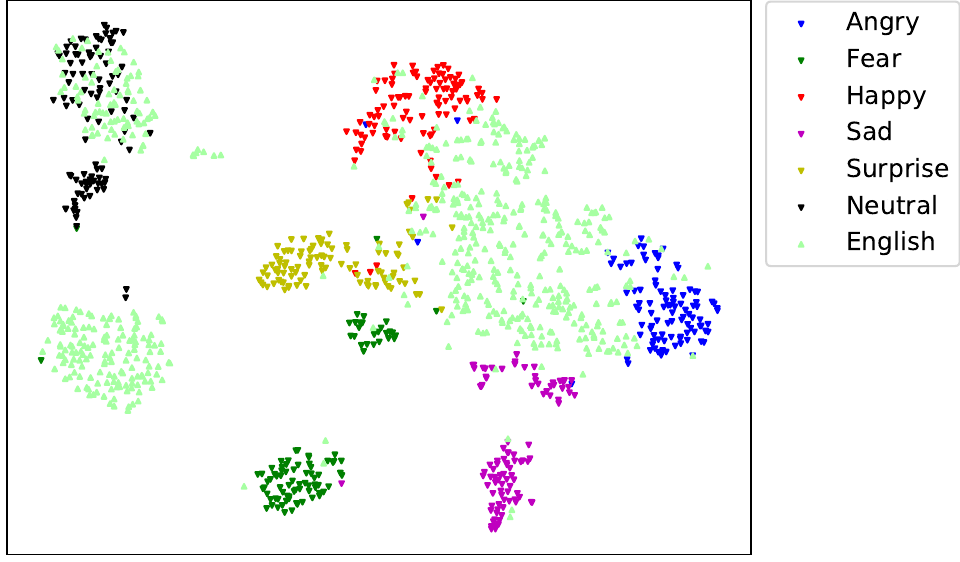}
\caption{Colored by emotion.}
\vspace{+10pt}
\end{subfigure}
\begin{subfigure}[b]{0.9\textwidth}
\centering
\includegraphics[width=\textwidth]{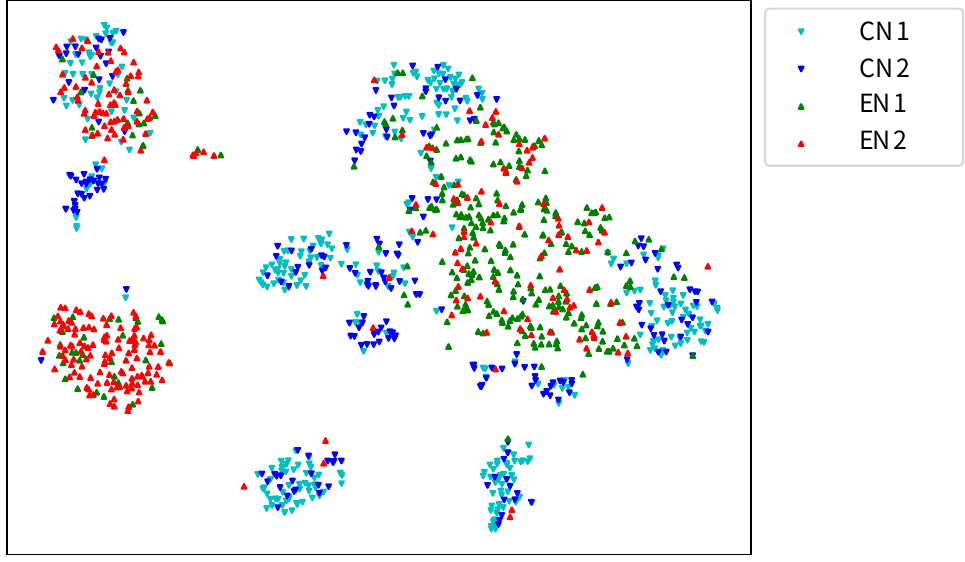}
\caption{Colored by speaker.}
\end{subfigure}
\end{minipage}
\caption{T-SNE visualization of emotion embedding. The difference between (a) and (b) is that they are colored by different attributes.}
\label{fig_3}
\end{figure}

We further visualize 
We further visualize the coarse-grained emotional representation via T-SNE~\cite{Maaten2008VisualizingDU}. Specifically, we preserve 100 utterances per emotion in the Chinese training speech data and 600 in English.

Figure~\ref{fig_3}(a) presents the T-SNE visualization of the emotion embeddings for Chinese utterances, demonstrating clear clusters. This observation validates the effectiveness of our semi-supervised emotion classifier. However, in Figure~\ref{fig_3}(a), we notice that certain emotion embeddings of English utterances are intermixed with those of Chinese utterances. We hypothesize that the language-agnostic nature of the coarse-grained emotion representation enables it to capture subtle emotional expressions in English utterances, resulting in their clustering alongside the Chinese utterances. This intermixed phenomenon of coarse-grained emotion representation signifies the METTS family's ability to transfer emotions across languages.

To further explore the extent to which the emotional representation encompasses the speaker's timbre attribute, we color the T-SNE visualization based on speaker ID. Figure~\ref{fig_3}(b) illustrates that the emotion embeddings are not well clustered according to the speaker, providing evidence of the speaker's independence in the coarse-grained emotion representation. This finding reinforces the effectiveness of our approach in disentangling speaker characteristics and isolating them from the emotional representation.


\section{Component analysis}
\label{component}

In Section~\ref{sc:results}, we demonstrate the excellent performance of METTS in both intra- and cross-lingual scenarios of emotional speech synthesis.
In this section, we aim to evaluate the effectiveness of each component by examining their impact on naturalness, speaker similarity, and emotion similarity.
Additionally, we analyze the influence of different values of clusters on the performance of METTS-ID.

\subsection{Ablation study of METTS-REF}

\begin{table*}[]
\centering
\caption{Results of different values of $N$ on model's performance with 95$\%$ confidence interval for Chinese speakers.}
\label{tab_11}
\begin{tabular}{@{}c|cccc|cccc@{}}
\toprule
   & \multicolumn{4}{c|}{Chinese Text}                                             & \multicolumn{4}{c}{English Text}                                              \\ \midrule
$N$  & Accuracy       & Naturalness        & Speaker Similarity & Emotion Similarity & Accuracy       & Naturalness        & Speaker Similarity & Emotion Similarity \\ \midrule
32 & \textbf{0.916} & 3.91±0.13          & 3.60±0.15          & 3.51±0.19          & \textbf{0.920} & 3.81±0.13          & 3.60±0.15          & 2.95±0.21          \\
64 & 0.854          & \textbf{4.07±0.13} & \textbf{3.88±0.16} & \textbf{3.95±0.13} & 0.875          & \textbf{4.06±0.18} & \textbf{3.77±0.20} & \textbf{3.24±0.22} \\
96 & 0.656          & 4.00±0.12          & 3.76±0.14          & 3.68±0.16          & 0.664          & 3.86±0.14          & 3.63±0.17          & 2.98±0.20          \\ \bottomrule
\end{tabular}
\end{table*}

\begin{table*}[]
\centering
\caption{Results of different values of $N$ on model's performance with 95$\%$ confidence interval for English speakers.}
\label{tab_12}
\begin{tabular}{@{}c|cccc|cccc@{}}
\toprule
   & \multicolumn{4}{c|}{Chinese Text}                                             & \multicolumn{4}{c}{English Text}                                              \\ \midrule
$N$  & Accuracy       & Naturalness        & Speaker Similarity & Emotion Similarity & Accuracy       & Naturalness        & Speaker Similarity & Emotion Similarity \\ \midrule
32 & \textbf{0.916} & 3.80±0.17          & 3.62±0.17          & 3.57±0.19          & \textbf{0.920} & 3.76±0.18          & 3.65±0.18          & 3.21±0.17          \\
64 & 0.854          & \textbf{4.02±0.15} & 3.57±0.21          & \textbf{3.73±0.17} & 0.875          & \textbf{4.05±0.18} & \textbf{3.74±0.18} & \textbf{3.26±0.14} \\
96 & 0.656          & 3.90±0.20          & \textbf{3.72±0.15} & 3.64±0.17          & 0.664          & 3.74±0.16          & 3.52±0.18          & 3.10±0.22          \\ \bottomrule
\end{tabular}
\end{table*}
We conduct ablation studies where the GST module, CVAE module, and perturb module are removed individually. The corresponding results are presented in Table~\ref{tab_9} and Table~\ref{tab_10}, respectively.

The removal of the GST module significantly affects the control of global emotional expression in bilingual speech. Without GST, METTS fails to map emotional expressions of different languages to the same global token and provide global emotion conditions. As a result, there is a significant decrease in emotion similarity and noticeable declines in naturalness and speaker similarity. This highlights the crucial role of the coarse-grained language-agnostic emotion representation in our approach.

Furthermore, when the CVAE module is removed, there is a sharp decline in naturalness and a decrease in emotion similarity. This indicates that the fine-grained emotional representation learned by the CVAE module, which is consistent with the input text, not only enhances the emotional expression but also plays a vital role in addressing the foreign accent problem and improving the overall naturalness of the synthetic speech.


Regarding the perturbation module, its omission slightly increased emotion similarity in most test conditions. However, it significantly compromised naturalness and speaker similarity. This trade-off suggests a substantial entanglement between speaker timbre, emotion, and language in multilingual emotional speech synthesis. Speaker timbre entangled with language may lead to abnormal pronunciation, while speaker timbre entangled with emotion may result in slightly high emotional expressiveness but low speaker similarity. Therefore, the necessity of speaker disentanglement becomes apparent to achieve idiomatic pronunciation and natural emotional expression for each speaker.

\subsection{Ablation study of METTS-ID}

Given the significance of the codebook size in Vector Quantization (VQ), we investigate the impact of different values of $N$ in the emotion matcher module on the performance of METTS-ID. Alongside evaluating naturalness, speaker similarity, and emotion similarity, we also examine the accuracy of the emotion matcher. For analysis, we retain 100 utterances per emotion in Chinese and 600 in English.

We first evaluate the accuracy of the emotion matcher by extracting the ground-truth cluster labels for each utterance and calculating the predicted accuracy. The results, presented in Table~\ref{tab_11} and Table~\ref{tab_12}, demonstrate that as the value of $N$ increases, there is an increase in the diversity of emotion embeddings. In contrast, the predicted accuracy of the emotion matcher gradually decreases. This indicates a complex trade-off between the diversity of emotion embeddings and the predicted accuracy of the emotion matcher. Notably, the predicted accuracy remains consistent across target speakers and languages, ensuring that METTS-ID can generate natural and emotionally expressive bilingual speech for each target speaker.

Furthermore, as shown in Table~\ref{tab_11} and Table~\ref{tab_12}, the effect of $N$ on speaker similarity is negligible, while naturalness and emotion similarity achieve their highest scores when $N$ is set to 64. Therefore, considering the overall performance, we designate $N$ as 64 to strike a balance between predicted accuracy, naturalness, speaker similarity, and emotion similarity.

\section{Conclusion}
\label{sc:conclusion}

This paper proposes METTS for multilingual emotional speech synthesis, aiming at achieving natural and diverse bilingual emotional speech across speakers. First, we introduce multi-scale emotion modeling to learn emotional expressions from a language-agnostic emotion representation (coarse-grained) and a language-specific emotion representation (fine-grained), effectively addressing the foreign accent problem. Meanwhile, we leverage information perturbation to address the problem of speaker timbre coupling and obtain speaker-independent multi-scale emotion representation. Moreover, we design a VQ-based emotion matcher to construct an embedding-candidate pool and select appropriate references according to the input text and emotion category for better emotional diversity and the naturalness of synthetic speech. English-Chinese bilingual experiments show that METTS can synthesize expressive bilingual speech with natural emotion and native pronunciation for each mono-lingual speaker.

During our investigation of the multilingual emotional TTS system through cross-speaker cross-lingual emotion transfer, we have identified the need for further improvements in synthesizing English emotional speech for both Chinese and English speakers. This is mainly because the English training corpus is mainly neutral in our study. We believe that leveraging emotional English corpus to train METTS will effectively improve the expressiveness of English synthetic speech in multilingual emotional text-to-speech.

\bibliographystyle{IEEEtran}
\bibliography{mybibfile.bib}
%




\end{document}